\documentclass[11pt,reqno]{article}

\usepackage{mathtools}
\usepackage{dirtytalk}
\usepackage{amsmath}
\usepackage{amsfonts}
\usepackage{amssymb}
\usepackage{amsxtra}
\usepackage{amsthm}
\usepackage{graphicx}
\usepackage{caption}
\usepackage{ushort}
\usepackage{color}
\usepackage{hyperref}
\usepackage[parsep]{collref}
\hypersetup{linktocpage=true}
\usepackage{enumitem}
\usepackage{titlesec}
\usepackage{cite}
\usepackage{slashed}
\usepackage{braket}
\usepackage{flexisym}
\usepackage{xcolor}
\usepackage{dsfont}
\usepackage{titlesec}
\usepackage{romannum}
\usepackage{caption}
\usepackage{subcaption}
\usepackage{mathrsfs}
\usepackage[bottom]{footmisc}

\def\ie{{\it i.e.\ }}

\newcommand{\mb}{\mathbb}

\newcommand{\ph}{\phantom}

\DeclareMathOperator{\vol}{vol}
\DeclareMathOperator{\Vol}{Vol}

\begin{document}
 
\pagenumbering{arabic} 

\noindent 
\LARGE

$^{}$
\vspace{1cm}
\begin{center}
\textbf{How to Create Universes with Internal Flux}
\end{center}
\normalsize

\vspace{1cm}
\begin{center}

Jean-Luc Lehners$^1$, Rahim Leung$^2$ and K.S. Stelle$^2$ \\
\vspace{0.4cm}

${}^1$ Max--Planck--Institute for Gravitational Physics (Albert--Einstein--Institute) \\ 14476  Potsdam, Germany \\

${}^2$ The Blackett Laboratory, Imperial College London \\ Prince Consort Road, London SW7 2AZ \\

\end{center}

\setlength\parindent{0pt}
\setcounter{footnote}{0}
\vspace{2cm}
\hrule
\vspace{.4cm}
String compactifications typically require fluxes, for example in order to stabilise moduli. Such fluxes, when they thread internal dimensions, are topological in nature and take on quantised values. This poses the puzzle as to how they could arise in the early universe, as they cannot be turned on incrementally. Working with string inspired models in $6$ and $8$ dimensions, we show that there exist no-boundary solutions in which internal fluxes are present from the creation of the universe onwards. The no-boundary proposal can thus explain the origin of fluxes in a Kaluza-Klein context. In fact, it acts as a selection principle since no-boundary solutions are only found to exist when the fluxes have the right magnitude to lead to an effective potential that is positive and flat enough for accelerated expansion. Within the range of selected fluxes, the no-boundary wave function assigns higher probability to smaller values of flux. Our models illustrate how cosmology can act as a filter on a landscape of possible higher-dimensional solutions. 
\vspace{.4cm}
\hrule
\newpage

\section{Introduction}

To date, string theory represents the most convincing way of combining quantum principles and gravity. This is because the quantisation of strings leads to remarkable consistency conditions, which not only require spacetime and matter to obey generalisations of Einstein's equations, but in fact also require the presence of additional spatial dimensions \cite{Polyakov:1981re}. Extra dimensions offer the prospect of unifying geometry and matter, as already exemplified in the original Kaluza-Klein model: in that model, gravity, electromagnetism and a scalar field in $4$ dimensions are seen to arise from a pure gravitational theory in $5$ dimensions compactified on a circle \cite{Kaluza:1921tu,Klein:1926tv}. \\

Additional spatial dimensions are, however, only compatible with observations if they lead to an effective $4-$dimensional spacetime at currently accessible energy scales, and moreover if the extra dimensions are static. The former criterion can be achieved either if the extra dimensions are compact and of sufficiently small volume \cite{Candelas:1985en} or if they are warped such that gravity localises on a $4-$dimensional submanifold \cite{Randall:1999vf,Crampton:2014hia}. The staticity criterion is necessary because time changing extra dimensions would lead to variations in coupling constants, most notably in the fine structure constant and in Newton's constant, and there exist stringent upper bounds on such variations \cite{Uzan:2002vq}. It is, however, difficult to keep the volume and shape of the extra dimensions static, especially when the uncompactified dimensions describe an expanding universe. This is the problem of moduli stabilisation. \\

String theory also contains form fields, which may be understood as generalisations of electromagnetism. A key realisation was that form fields may wrap the extra dimensions (often in combination with branes), thereby producing potentials for the moduli \cite{Sethi:1996es,Giddings:2001yu}. When such potentials contain minima, the corresponding moduli can be stabilised. Fluxes, seen as possible sources of branes, obey generalised Dirac quantisation conditions. They are naturally proportional to integer values, and cannot be turned on incrementally. This raises the question as to how higher-dimensional spacetimes with internal fluxes could have arisen in the first place \footnote{Once flux is present, its amount can be changed by nucleations of membranes \cite{Brown:1988kg}. Such tunnelling effects take place in existing universes, while we are asking how such universes could have come into existence.}. Moreover, the potentials induced by fluxes depend rather crucially on the amount of flux -- if there is too little or too much flux, the potentials typically do not lead to the desired stabilisation of moduli. Hence a second question is how the appropriate amount of flux may be fixed. Our aim in the present paper is to provide a possible answer to both riddles. \\

Our answer involves the no-boundary proposal \cite{Hartle:1983ai}, which is a theory of initial conditions for the universe. Though originally formulated in $4$ dimensions, we find that it leads to interesting consequences in the higher-dimensional context. The no-boundary proposal is most easily formulated in semi-classical gravity, specifically in the path integral formulation. The idea is to calculate the wave function of the universe by summing over geometries and matter configurations that are compact and regular. As we will review, this is possible only if one allows for complex saddle points of the path integral. Thus the creation of the universe may be seen as a quantum process, a process that would have been forbidden in classical physics. \\

The shape of no-boundary solutions depends on the theory under consideration. We will consider toy models, but using only ingredients that are naturally present in string theory. Our main example involves an $8-$dimensional theory of gravity and $4-$form flux, where in addition we consider the leading $\alpha^\prime$ correction, namely a $R^4$ correction term. This term acts as an effective source of vacuum energy \cite{Ketov:2017aau,Otero:2017thw}, just as in the Starobinsky model of inflation in $4$ dimensions \cite{Starobinsky:1980te}. The theory is then compactified on a $4-$sphere. With the appropriate amount of flux, this model induces an effective $4-$dimensional scalar potential that contains a local minimum for the radion field (which determines the size of the internal sphere). In addition, it contains an inflationary plateau for the scalaron field. The inflationary plateau plays a crucial role, as it gives rise to a dynamical attractor. This is necessary in order for no-boundary solutions to exist, because these solutions must interpolate between no-boundary initial conditions and real final field values. It is only when a dynamical attractor is present (provided by inflation here) that such an interpolation becomes possible \cite{Hartle:2008ng,Lehners:2015sia}. And an inflationary potential only arises for a certain range of fluxes, implying that the no-boundary proposal automatically selects an amount of flux in that range. Otherwise, one simply does not obtain a large, classical universe. Moreover, within the allowed range, the no-boundary wave function assigns probabilities to different amounts of flux, typically favouring flux amounts that lead to a lower potential. In this way, the no-boundary proposal provides a probability measure on the landscape of stringy solutions. An interesting questions is then whether this measure gives reasonable results. This question cannot be fully answered at present. On the one hand, the no-boundary measure helps in explaining why moduli fields would be stabilised at minima of their potential, but on the other hand it does not fully answer the question of why inflation lasted for many e-folds, as a few e-folds would have been enough to create a classical universe. More complete models might eventually clarify this issue.\\

A second model that we discuss is inspired by the embedding of Salam-Sezgin theory in string theory. This $6-$dimensional model is distinguished by its containing a positive potential \cite{Salam:1984cj}, which however is not quite steep enough to allow for inflationary solutions (in fact it sits just at the boundary of inducing accelerated or decelerated expansion). We consider a generalisation of the model, with a slightly flatter potential, compactified on a $2-$sphere with $2-$form flux. Once again we find no-boundary solutions, describing the nucleation of large, classical universes. A caveat of this model is that the internal $2-$sphere eventually decompactifies, as the potential does not develop a local minimum for any amount of flux. From a Kaluza-Klein point of view, this model is of particular interest \cite{Cvetic:2003xr}\cite{Crampton:2014hia}, but further ingredients will have to be included to render it more realistic. \\

Our paper is organised as follows: we start with a brief review of the no-boundary proposal. Then we will investigate no-boundary solutions in an $8-$dimensional model in Section \ref{sec:8d}, and in the generalised Salam-Sezgin model in Section \ref{sec:6d}. Our conclusions are presented in Section \ref{sec:discussion}, and an appendix provides further details regarding the equations of motion of our models. 


\section{The no-boundary proposal} \label{sec:noboundary}

The no-boundary proposal is a (quantum) theory of boundary conditions for the universe, most transparently formulated in the language of path integral quantisation \cite{Hartle:1983ai} (see also \cite{Hartle:2008ng} for an overview, and \cite{Feldbrugge:2017kzv,DiazDorronsoro:2017hti,Halliwell:2018ejl,Lehners:2021jmv} for more recent discussions). One starts by considering the wave function of the universe, written as a sum over geometries $g_{\mu\nu}$ and matter configurations $\phi$ interpolating between initial and final conditions,
\begin{align}
\Psi(g_{1ij},\phi_1) = \int {\cal D} g_{\mu\nu} {\cal D} \phi e^{\frac{i}{\hbar}S} = \int {\cal D} g_{\mu\nu} {\cal D} \phi e^{\frac{i}{\hbar}\int_{n.b.}^{g_{ij}=g_{1ij},\, \phi=\phi_1}d^Dx \sqrt{-g}{\cal L}}\,, \label{wavefunction}
\end{align}
where $S$ denotes the action and ${\cal L}$ the corresponding Lagrangian. Here we have denoted the final $(D-1)-$dimensional hypersurface by a subscript $1,$ and have demanded that the metric and fields take the values $g_{1ij},\phi_1$ there. The no-boundary initial conditions, denoted $n.b.$ in the equation above, are chosen such that the saddle points of the path integral are formed by geometries that contain no boundary in the past, \ie such that the geometries are compact and regular. The matter configurations on these saddle point geometries must also be regular. The prototype of such a geometry occurs when considering the theory of gravity coupled to a positive cosmological constant $\Lambda=3H^2$ with Lagrangian density ${\cal L} = R/2 - \Lambda,$ where $R$ denotes the Ricci scalar curvature. Then a saddle point of the path integral (satisfying $\delta S =0$) consists of a solution to the equations of motion -- in this case the solution is de Sitter spacetime. The no-boundary condition can be satisfied by considering a section of the complexified de Sitter solution. In $4$ dimensions the de Sitter solution can be written explicitly as
\begin{align}
ds^2 = - N^2 dt^2 + \frac{1}{H^2}\cosh^2(HNt) d\Omega_3^2\,,
\end{align}
where $d\Omega_3^2$ is the metric on a round 3-sphere, and the lapse is trivial, $N=1$. de Sitter space can be understood as a hyperboloid embedded in $5$ dimensions.  The realisation of Hartle and Hawking was that at the waist ($t=0$) of the hyperboloid, we can glue on a Euclidean section of de Sitter spacetime, which corresponds simply to half of a $4-$sphere. The necessary continuation is
\begin{align}
t= - i \left(\tau - \frac{\pi}{2H}\right)\,, \qquad \frac{\pi}{2H} \geq \tau \geq 0 \,,\label{ancon}
\end{align}
with the metric along the Euclidean section being
\begin{align}
ds^2 = d\tau^2 + \frac{1}{H^2}\sin^2(H\tau) d\Omega_3^2\,.
\end{align}
This has the consequence that now the geometry rounds off smoothly at $\tau=0,$ which is sometimes referred to as the South Pole of the geometry. In general one could consider a different path in complexified time, in particular a smooth path, interpolating between $\tau=0$ and the final time on which the desired final values of the fields are reached. If there are no singularities, then Cauchy's theorem implies that the action remains unchanged -- geometries related by such a change of path (with end points fixed) yield the same action and should be regarded as equivalent. We will refer to the choice made above, purely Euclidean followed by purely Lorentzian, as a ``Hawking'' contour \cite{Hawking:1981gb}. \\

A few remarks are in order: First, since there is no initial boundary, the wave function \eqref{wavefunction} is a function of the final values of the fields only. Mathematically implementing the no-boundary condition can however be tricky, as compactness and regularity are typically conditions that cannot be imposed simultaneously, due to the Heisenberg uncertainty relations (compactness is a condition on the spatial volume, while regularity is a condition on the expansion rate, which is canonically conjugate to the volume) \cite{DiTucci:2019dji}; in simple models, it has proven fruitful to impose just the regularity condition, see \cite{DiTucci:2019bui,Lehners:2021jmv}\footnote{This choice is also corroborated by an analogy with AdS/CFT \cite{Caputa:2018asc,DiTucci:2020weq}.}. Since we will be concerned purely with the saddle point geometries, this subtlety will not be of importance for us. \\

Second, the action picks up an imaginary contribution due to the Euclidean part of the geometry, specifically the integral along the $4-$sphere part of the geometry gives  $-i\frac{4\pi^2}{H^2}.$ This means that the wave function obtains a weighting
\begin{align}
|\Psi| \, \approx \, e^{\frac{12\pi^2}{\hbar \Lambda}}\,.
\end{align}
If one allows $\Lambda$ to change or, rather, if one generalises the model by adding a scalar field with a potential $V(\phi)$, then this relation generalises to \cite{Hartle:2008ng}
\begin{align}
|\Psi| \, \approx \, e^{\frac{12\pi^2}{\hbar \, |V(\phi_{SP})|}}\,,
\end{align}
where $\phi_{SP}$ denotes the (generally complex) value of the scalar field at the South Pole of the geometry. This formula implies that lower initial values of the potential come out as preferred.  In this way the no-boundary proposal provides probabilities for different histories. We should add that if the complex conjugate choice of Wick rotation had been made in \eqref{ancon}, then the weighting would have been the inverse, and higher potential values would have been preferred \cite{Vilenkin:1983xq}. However, such a choice yields an inconsistent model, in which tensor fluctuations obey an inverse Gaussian distribution \cite{Halliwell:1984eu,Halliwell:1989dy,Feldbrugge:2017fcc}, and thus the sign of the Wick rotation is fixed to be that in \eqref{ancon}. \\

Third, the no-boundary geometry reaches zero size yet avoids a big bang type singularity. This is made possible precisely by allowing the metric (and generally matter fields) to take complex values. This is very much in analogy with ordinary tunnelling in quantum mechanics, where classically impossible boundary conditions can be overcome by allowing fields (or time, equivalently the lapse function) to become complex \cite{Turok:2013dfa,Cherman:2014sba,Bramberger:2016yog}. However, an important requirement for no-boundary solutions to exist is that they must reach the desired final, real field values imposed on the final hypersurface. These final values, which are the arguments of the wave function, are the physically measurable quantities, and thus must be real valued. Reaching real values is, however, a non-trivial requirement, and in fact only occurs when the dynamical theory contains an attractor. Just two cosmological examples are currently known: inflation \cite{Hartle:2008ng} and ekpyrosis \cite{Battarra:2014kga,Lehners:2015sia}. We will focus on inflationary solutions, as ekpyrotic universes require an ill-understood bounce phase to reach the current expanding phase of the universe. The attractor has a further consequence: it drives the wave function to a semi-classical (WKB) form in which the phase of the wave function changes rapidly as the universe expands, while the weighting tends to a constant value. This feature allows one to associate probabilities to different cosmological histories.\\

After this brief overview, we are in a position to look for no-boundary solutions in a higher-dimensional setting. We will not consider off-shell configurations in the path integral \eqref{wavefunction}, nor issues of integration contours, leaving these questions for future work. Rather, we will focus on saddle point solutions which we expect to provide the dominant contributions to the no-boundary wave function.


\section{An $8$-dimensional Starobinsky model} \label{sec:8d}

Low-energy approximations to string theory, to order ${\cal O}(\alpha^{\prime 0}=1)$ in the string length, do not seem to allow for de Sitter solutions \cite{Obied:2018sgi}. In fact, obtaining even a short, transient,  period of accelerated expansion proves difficult \cite{Agrawal:2018own}. However, both due to the extended nature of the strings as well as due to quantum corrections, we expect higher curvature corrections to be present. These can act as effective potentials, opening the possibility of accelerated solutions. Here we will consider the case where an $R^m$ term is added to the higher-dimensional action \cite{Ketov:2017aau}. Our aim is not to find an explicit embedding into string theory (though this is a highly desirable goal for future work), but rather to consider an example where a higher curvature term proves useful in obtaining solutions of potential cosmological relevance. \\

Thus we take the starting action in $D$ dimensions to be given by
\begin{align}
    S &= \frac{1}{2}\int d^{D} x \sqrt{-\hat{g}}\left(\hat{R}+ \alpha \hat{R}^m - \frac{1}{2p!}q^2 F_{(p)}^2\right) \,,
\end{align}
with $m$ a positive integer and we consider a generic $p$-form flux $F_{(p)}$ with coupling $q,$ satisfying a Bianchi identity $dF_{(p)} = 0$. As is well known, the $R^m$ term effectively introduces a new scalar degree of freedom \cite{Stelle:1977ry,Whitt:1984pd}. This can be made manifest by performing a conformal transformation on the metric,
\begin{align}\label{conformal}
    \hat{g}_{\mu\nu} \equiv e^{2\varphi} g_{\mu\nu}\,,
\end{align}
under which 
\begin{align}
    \sqrt{-\hat{g}} = e^{D\varphi} \sqrt{-g}\,, \qquad \hat{R} = e^{-2\varphi}\left[R - 2(D-1)\Box\varphi - (D-1)(D-2)\nabla^\mu \varphi \nabla_\mu \varphi \right]\,.
\end{align}
The trick now is to rewrite the action as
\begin{align}
    S = \frac{1}{2}\int d^D x \sqrt{-\hat{g}} \left( f_{,\hat{R}} \hat{R} - U - \frac{1}{2p!}q^2 F_{(p)}^2\right) 
\end{align}
with $U = (f_{,\hat{R}}\hat{R}-f)$ and $f=\hat{R}+ \alpha \hat{R}^m.$ Then we obtain
\begin{align}
    S = \frac{1}{2}\int d^D x \sqrt{-g} & \left(  f_{,\hat{R}} e^{(D-2)\varphi}\left[ R - 2(D-1)\Box\varphi - (D-1)(D-2)\nabla^\mu \varphi \nabla_\mu \varphi \right]  \nonumber \right. \\ & \left. - e^{D\varphi} U - \frac{1}{2p!} e^{(D-2p)\varphi} q^2 F_{(p)}^2 \right) \,. \label{eq1} 
\end{align}
As long as we choose
\begin{align}
e^{(2-D)\varphi} = f_{,\hat{R}} = (1 + m\alpha \hat{R}^{m-1}) \,,
\end{align}
we will end up in Einstein frame. Here we must assume that $1+m\alpha \hat{R}^{m-1} >0.$ This choice has the added benefit that the $\Box \varphi$ term in \eqref{eq1} turns into a total derivative, and can be dropped. The potential is then given by
\begin{align}
    V(\varphi) = \frac{1}{2}e^{D\varphi} U =  \frac{\alpha(m-1)}{2(m\alpha)^{\frac{m}{m-1}}}\left( e^{(2-D+\frac{m-1}{m}D)\varphi} - e^{\frac{(m-1)D}{m}\varphi}\right)^{\frac{m}{m-1}}
\end{align}
Now we can see that the potential develops a plateau, \ie a region of the potential which is positive and very flat, when $2-D+\frac{m-1}{m}D=0,$ \ie when $D=2m$ \cite{Ketov:2017aau}. This occurs for instance in Starobinsky's model of inflation, in $D=4$ and with an $\hat{R}^2$ term in the action \cite{Starobinsky:1980te}. Below, we will consider the example of $D=8$ and consequently fix $m=4.$ \\

As a final step, we can rescale the scalar to make it canonically normalised,
\begin{align}
    \sqrt{(D-1)(D-2)}\varphi \equiv \phi\,,
\end{align}
so that the action ends up being given by
\begin{align}
    S & = \frac{1}{2}\int d^{D} x \sqrt{-\hat{g}}\left(\hat{R}+ \alpha \hat{R}^4 - \frac{1}{2}q^2 F_{(p)}^2\right) \\ & = \int d^D x \sqrt{-g} \left( \frac{R}{2} - \frac{1}{2}\nabla^\mu \phi \nabla_\mu \phi - V(\phi) - \frac{1}{4p!}q^2 e^{\frac{D-2p}{\sqrt{(D-1)(D-2)}}\phi} F_{(p)}^2\right)
\end{align}
with potential 
\begin{align}
    V(\phi) = \frac{3}{2^{11/3}\alpha^{1/3}}\left(e^{(\frac{D}{4}-2)\varphi} - e^{-\frac{3D}{4}\varphi} \right)^{\frac{4}{3}} = \tilde\alpha\left(1 - e^{-\sqrt{\frac{6}{7}}\phi} \right)^{\frac{4}{3}}\,,
\end{align}
where we inserted $D=8$ into the last expression and redefined the numerical prefactor for convenience. \\

In the lower-dimensional theory, there will be an extra contribution to the potential coming from the flux term. As we discussed in Section \ref{sec:noboundary}, in order for no-boundary solutions to exist, an attractor must be present. The best known example of a cosmological attractor is an inflationary one, and we will focus on this example. However, inflation requires a suitably flat potential, and the generically rather steep $\phi-$dependence resulting from the flux term makes it difficult to find potentials that are flat enough. There is one exception \cite{Otero:2017thw}, which occurs when $p=D/2,$ as in that case the $\phi-$coupling to the flux disappears. Thus we will choose $p=4.$ The equations of motion are presented in equations \eqref{8dscalar}-\eqref{8dmetric} in Appendix \ref{apA}. As an aside, we note that terms quartic in the Riemann curvature tensor, as well as a $4-$form gauge potential, arise in 11-dimensional supergravity \cite{Cremmer:1978km,Deser:1977yyz,Green:1997as}, and will thus be present generically upon compactification down to $8$ dimensions. \\

The metric ansatz for our compactification reads
\begin{align}\label{8dansatz2}
    ds_8^2 = e^{-\frac{2}{\sqrt{3}}\chi}ds_4^2 + e^{\frac{1}{\sqrt{3}}\chi}d\Omega^{2}_4  \,.
\end{align}
Here $\chi$ parameterises the size of an internal 4-sphere. The numerical coefficients have been chosen so that the kinetic term for $\chi$ assumes canonical form in the dimensionally reduced theory. We will restrict to a 4-dimensional closed FLRW metric $ds_4^2 = -N^2 dt^2 + a(t)^2 d\Omega_3^2$ and correspondingly assume that the radion $\chi(t)$ depends only on time. We will also assume the presence of $4-$form flux on the sphere, with magnetic flux configuration 
\begin{align}
    F_{(4)} = 2 n_4 \vol(S^4)\,.
\end{align}
Here $n_4$ will be proportional to an integer flux quantum number \cite{Henneaux:1986ht} (the factor of $2$ is added for convenience), 
\begin{align}
    n_4 = \frac{2\pi}{2 q \vol(S^4)}z = \frac{3}{8\pi q} z\,, \qquad z \in \mathbb{Z}\,.
\end{align}
Note that the numerical values of $n_4$ are typically not integral, since the spacings between values are determined by the value of the charge $q.$ \\

The action then reads
\begin{align}
    S   &=  \frac{16\pi^4}{3} \int dt  \left(-3\frac{a \dot{a}^2}{N}+\frac{a^3}{2N}\left( \dot\phi^2 + \dot{\chi}^2 \right) +3Na - N a^3 V(\phi,\chi)\right) \nonumber \\ & \quad + \left[3\frac{a^2\dot{a}}{N} -\frac{a^3\dot{\chi}}{\sqrt{6}N} \right]_{surface}\,.
\end{align}
The surface terms can be eliminated on the final hypersurface by adding a York-Gibbons-Hawking boundary term there \cite{York:1972sj,Gibbons:1976ue}. On the initial hypersurface we will not add any boundary term (since there is not meant to be a boundary there), but they would vanish in any case for saddle point geometries with $a(0)=0.$ Effectively the action is that of a 4-dimensional scale factor $a(t)$ coupled to two scalar fields $\phi,\chi$ moving in an effective potential 
\begin{align}
    V(\phi,\chi) &= \tilde\alpha\left(1 - e^{-\sqrt{\frac{6}{7}}\phi} \right)^{\frac{4}{3}} e^{- \frac{2}{\sqrt{3}}\chi}  + n_4^2 e^{-2\sqrt{3}\chi} - 6 e^{-\sqrt{3}\chi}\,. \label{R4F4potential}
\end{align}

\begin{figure}[h]
	\centering
	\includegraphics[width=0.45\textwidth]{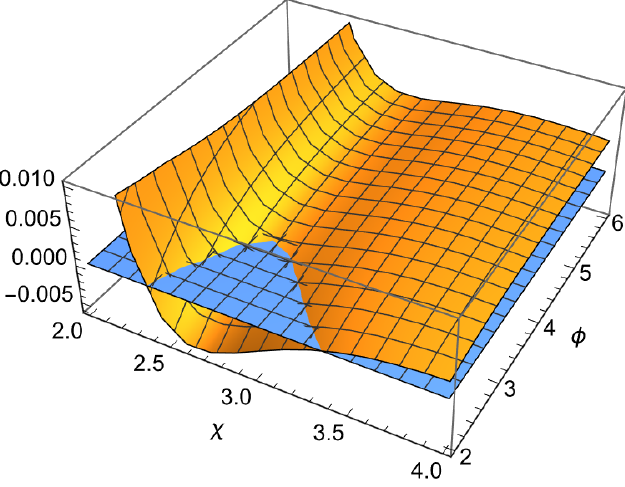}
	\includegraphics[width=0.45\textwidth]{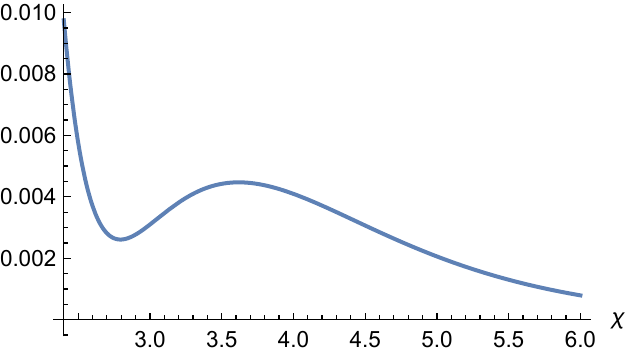}
	\caption{Potential for $\tilde\alpha=1$ and $n_4=13.$ The left panel contains a blue surface at $V=0$ for reference. The right panel provides a slice at $\phi=6,$ where we can see that a valley exists allowing $\chi$ to be stabilised at $\chi_{min} \approx 2.8$ (with a small slope along the orthogonal $\phi$ direction). The inflationary valley eventually drops to negative values, as is evident from the left panel.}
	\label{fig:8dpot}
\end{figure}

The potential contains a plateau in $\phi,$ in direct analogy with the Starobinsky model of inflation. There is a second field however, namely the radius of the internal sphere. For sufficiently large values of the flux $n_4$, this field develops a minimum at positive values of the potential, see the example in  Fig.\ \ref{fig:8dpot} where we chose $\tilde\alpha=1,\, n_4=13$. In such a case, a valley forms along which inflationary solutions can be expected to be found.\\

We can now look for examples of solutions, to see if this expectation is borne out.  The equations of motion are given by
\begin{align}
    0 & = 3 \frac{\ddot{a}}{a} +\dot\phi^2 + \dot{\chi}^2  - N^2 V\,, \\
    0 & = \ddot\phi + 3 H \dot\phi + N^2 V_{,\phi}\,, \\
    0 & = \ddot\chi + 3 H \dot\chi + N^2 V_{,\chi}\,,
\end{align}
while the constraint is
\begin{align}
    3 (\dot{a}^2 + N^2) = \frac{a^2}{2}(\dot\phi^2 + \dot\chi^2) + a^2 N^2 V\,.
\end{align}
The on-shell Lagrangian is obtained by using the constraint
\begin{align}
    S^{on-shell} 
    &= \frac{32\pi^4}{3} \int dt  \left(3Na - N a^3 V(\phi, \chi)\right)\,.
\end{align}

For $\phi \gtrapprox 1,$ the $\phi$ dependence in the potential \eqref{R4F4potential} becomes negligible in determining the minimum $\chi_{min},$ which is specified by a solution to the condition \cite{Otero:2017thw}
\begin{align}
    3 e^{\sqrt{3}\chi_{min}} -\frac{\tilde\alpha}{3} e^{ \frac{4}{3}\sqrt{3}\chi_{min}}  = n_4^2\,.  
\end{align}
At the minimum, the value of the potential is 
\begin{align}
    V_{min} = \frac{2\tilde\alpha}{3}e^{ -\frac{2}{3}\sqrt{3}\chi_{min}} -3 e^{ -\sqrt{3}\chi_{min}} \,.
\end{align}
Larger flux implies a larger $\chi_{min},$ which translates into a higher valley floor. Given that the location of the minimum in $\chi$ is approximately independent of $\phi$, this implies that at $\chi_{min}$ we obtain an effective potential that depends to good approximation solely on $\phi,$
\begin{align}
    V_{eff}(\phi) =V(\phi,\chi_{min}) \approx V_{min} \left(1 - e^{-\sqrt{\frac{6}{7}}\phi} \right)^{\frac{4}{3}}\,.
\end{align}

\begin{figure}[h]
	\centering
	\includegraphics[width=0.7\textwidth]{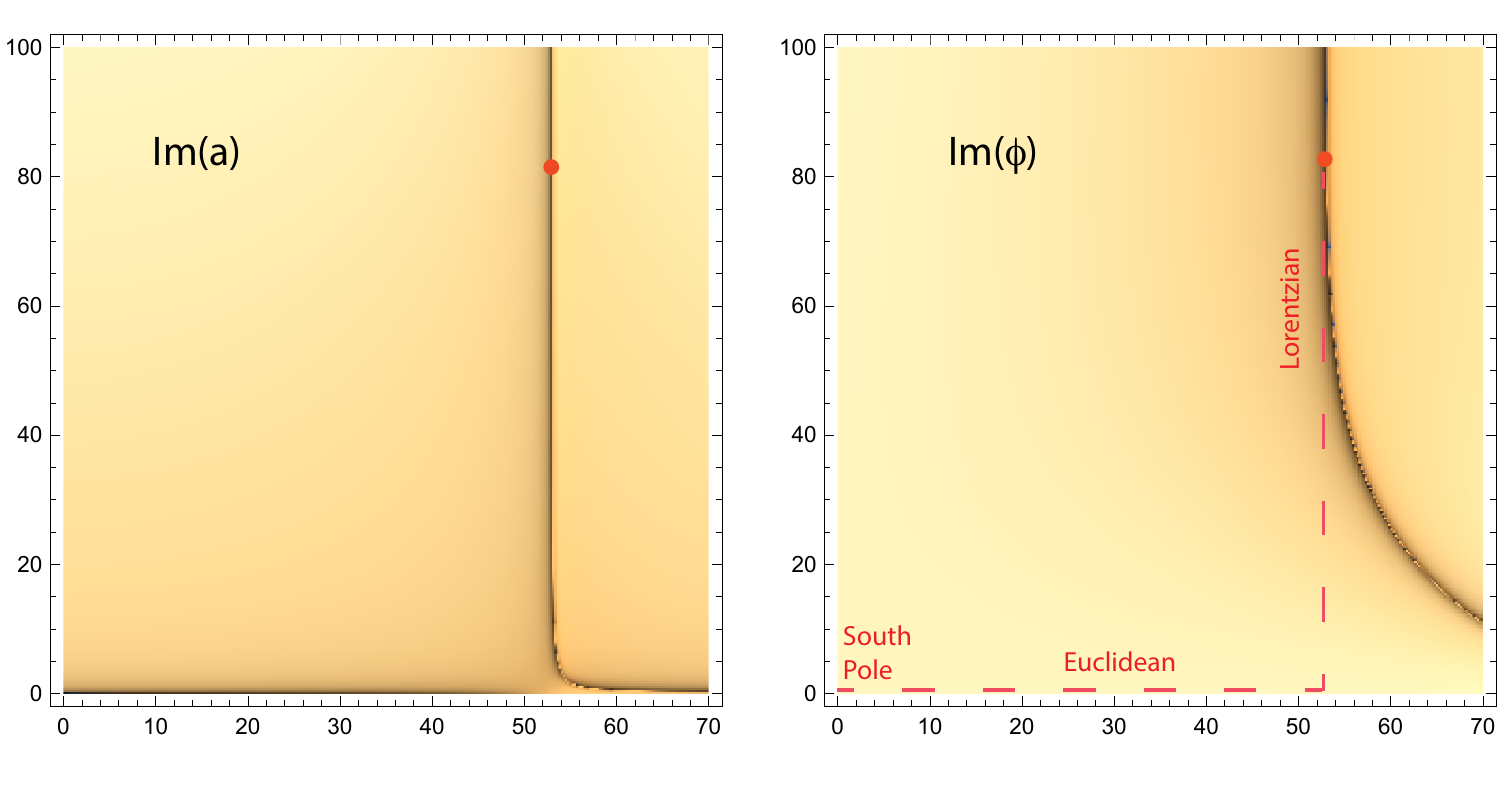}
	\caption{Density plots of the imaginary values of the scale factor and the scalar field, in the complex time plane. The South Pole resides at the origin; the horizontal axis corresponds to Euclidean time, the vertical axis to Lorentzian time. Darker colours indicate smaller imaginary values, so that the black lines show the locus of real field values. The dashed line in the right panel indicates the ``Hawking'' contour used in Fig.\ \ref{fig:ex8d2_2}. One can see that at late times, overlapping dark lines emerge in both plots, indicating that one approaches a real, classical solution of the equations of motion. The parameters chosen are  $\tilde\alpha=1, n_4=13.$ The present solution has constant sphere size $\chi_{min}=0.00281,$ initial scalar field value $\phi_{SP}=6.1104-0.09991i$ and it reaches the final values $a_1=200,\phi_1=6$ at time $\tau=53.185+83.538i,$ as marked by a red dot.}
	\label{fig:ex8d2_1}
\end{figure}

We are now looking for no-boundary solutions in this potential, \ie solutions that are regular and compact \cite{Jonas:2020pos}. Compactness means that we would like to set $a(0)=0.$ Regularity is best determined by expanding the equations of motion and the constraint as a Taylor series around the origin.  This is not complicated numerically, though the expressions beyond the leading terms can be rather lengthy. In the present case, the leading order terms (in Euclidean time) are found to be
\begin{align}
    a(\tau) &= \tau -\frac{1}{18} V(\phi_{SP}) \tau^3  + {\cal O}(\tau^5)\,, \\ 
    \phi(\tau) &=\phi_{SP}  +\frac{1}{8} V'(\phi_{SP}) \tau^2 + {\cal O}(\tau^4)\,.
\end{align}
Note that $\phi_{SP}$ is a complex integration constant, which we need to determine. It has to be adjusted such that $a$ and $\phi$ take specified (real) final values $a_1, \phi_1$ at some common final time $t_1.$ This is where the inflationary attractor is required: although it is not difficult to choose $\phi_{SP}$ such that the values $a_1,\phi_1$ are reached somewhere, the non-trivial part is that they must be reached simultaneously. We find $\phi_{SP}$ by using a numerical optimisation algorithm. More specifically, we integrate the equations of motion up to a point in the complexified time plane where the scalar field takes the specified value $\phi_1.$ Then we use a Newtonian algorithm to adjust the final time as well as $\phi_{SP}$ such that the scale factor $a$ approaches the desired value $a_1$ simultaneously up to a specified accuracy (which we take to be $8$ significant digits). \\

\begin{figure}[h]
	\centering
	\includegraphics[width=0.35\textwidth]{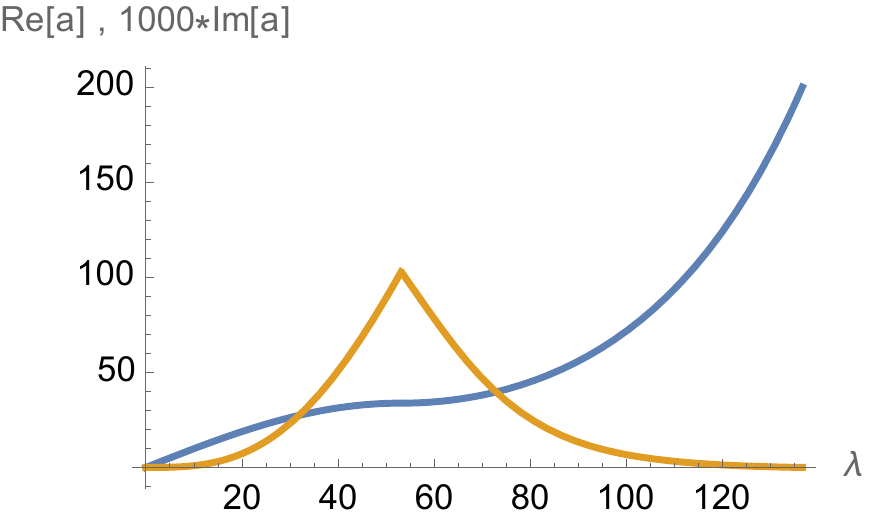}
	\includegraphics[width=0.35\textwidth]{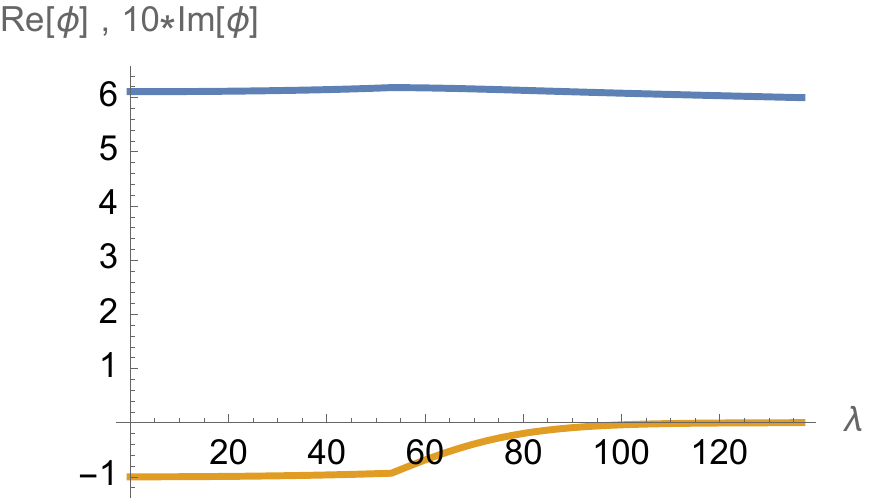}
	\caption{Field evolutions for the no-boundary solution shown in Fig.\ \ref{fig:ex8d2_1}, along a ``Hawking'' path, starting at the South Pole in a Euclidean direction and then turning 90 degrees into the Lorentzian time direction at the location where real field values are reached at the final time. Blue curves show the real parts, while orange curves show the imaginary parts (magnified $1000$ times for the scale factor, and $10$ times for the scalar field).}
	\label{fig:ex8d2_2}
\end{figure}

A representative example of a solution is provided in Figs.\ \ref{fig:ex8d2_1} and \ref{fig:ex8d2_2}. Fig.\ \ref{fig:ex8d2_1} provides a ``global'' view of the solution, as it shows a density plot of the imaginary parts of both the scale factor and the scalar field over the relevant region in the complexified time plane. A darker colour means a smaller imaginary part, so that the black lines in the figure show the locus of real field values. One can see that the scale factor is approximately real on a line segment protruding in the Euclidean time direction from the origin, and also on a vertical/Lorentzian line. The scalar field becomes real at large times on the exact same vertical/Lorentzian line. This shows that, despite the complex (quantum) starting point, the solution approaches a real, classical solution of the equations of motion at late times. In the right panel of Fig.\ \ref{fig:ex8d2_1} the dashed line shows the contour used to describe the typical de Sitter instanton in Section \ref{sec:noboundary}, following the Euclidean direction from the South Pole, and then turning abruptly into the Lorentzian direction so as to reach the final asymptotic classical solution. The field evolutions along this contour are shown in Fig.\ \ref{fig:ex8d2_2}. Note that we enhanced the imaginary parts of the fields for better visibility. One can see that the scalar $\phi$ is almost real at the South Pole, and then becomes exactly real on the final hypersurface. The scale factor is that of an approximate $4-$sphere morphing into a Lorentzian de Sitter solution. Thus these solutions can be regarded as small complex deformations of the de Sitter prototype solution. \\

It is important to recall that in addition to the fields shown here, the radion sits at its minimum, $\chi=\chi_{min}.$ This implies that these solutions describe a $4-$dimensional inflationary universe, with a stable internal $4-$sphere, and on that $4-$sphere we have $n_4$ units of magnetic flux. Thus this solution describes the creation of an $8-$dimensional universe, with precisely the Kaluza-Klein characteristics we searched for. We should point out that at the creation of the universe, \ie at $\tau=0,$ the scale factor is zero while the radius of the internal sphere already has size $\chi_{min}.$ Because this is a product geometry, the spatial volume of the universe is nevertheless zero there, as $\sqrt{-g} = a^3\Vol(S^4).$ This is very much in analogy with the Nariai no-boundary solutions with topology $dS^2 \times S^2$ discussed in \cite{Bousso:1995cc}, while here we have $dS^4 \times S^4$. \\

\begin{figure}[h]
	\centering
	\includegraphics[width=0.3\textwidth]{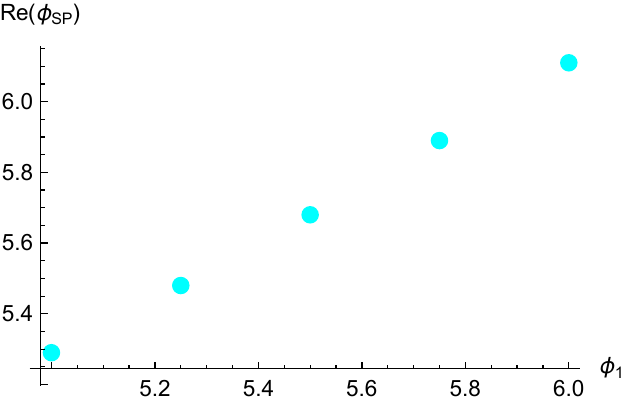}\,
	\includegraphics[width=0.3\textwidth]{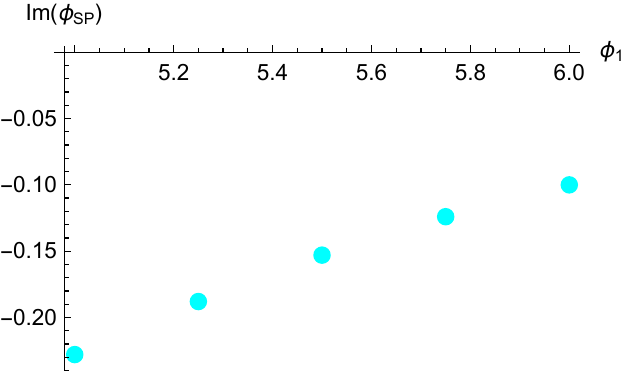}\,
	\includegraphics[width=0.3\textwidth]{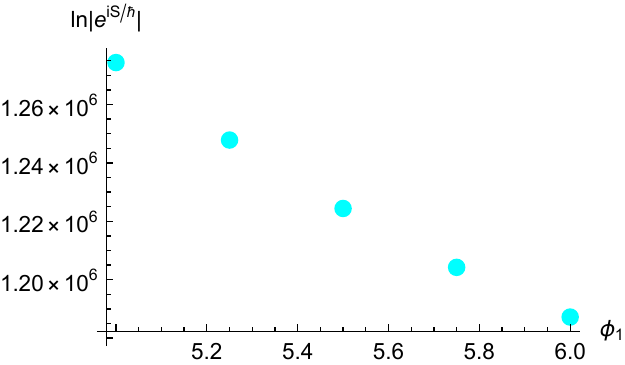}
	\caption{Optimised South Pole values and weightings for different final scalar field values $\phi_1,$ for fixed final scale factor $a_1=200.$ The parameters chosen are  again $\tilde\alpha=1, n_4=13.$ Solutions that emerge lower on the potential (smaller $\phi_1$) are seen to have higher probability, as is expected for no-boundary initial conditions.}
	\label{fig:series}
\end{figure}

Similar solutions can be found for other final values of the scalar field and scale factor. For fixed final scale factor $a_1=200,$ we show the corresponding South Pole values of the scalar field and the corresponding weightings in Fig.\ \ref{fig:series}. One can think of each such solution as being part of a series in which the fields continue to roll down the potential along a classical solution of the equations of motion, with the scale factor undergoing accelerated expansion in $4$ dimensions, and with the internal (flux-threaded) $4-$sphere remaining of fixed size. The probabilities for the different classical solutions are then provided by the weightings shown in Fig.\ \ref{fig:series}. As expected, the weightings are larger for solutions that emerge lower on the potential.

For small enough $\phi$ the potential turns negative, and thus the fate of universes such as those just described would be to collapse eventually. In this respect the solutions described here are just toy model solutions, lacking a proper graceful exit mechanism, though they demonstrate explicitly that universes with stable internal dimensions can be created via no-boundary instantons. \\

\begin{figure}[h]
	\centering
	\includegraphics[width=0.6\textwidth]{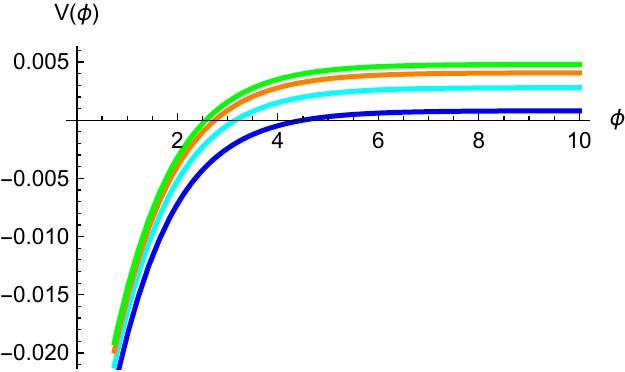}
	\caption{The effective potentials that arise in stable valleys for the fluxes $n_4=12,13,14,15.$ Larger flux corresponds to a higher potential.}
	\label{fig:ex8potentials}
\end{figure}

\begin{figure}[h]
	\centering
	\includegraphics[width=0.3\textwidth]{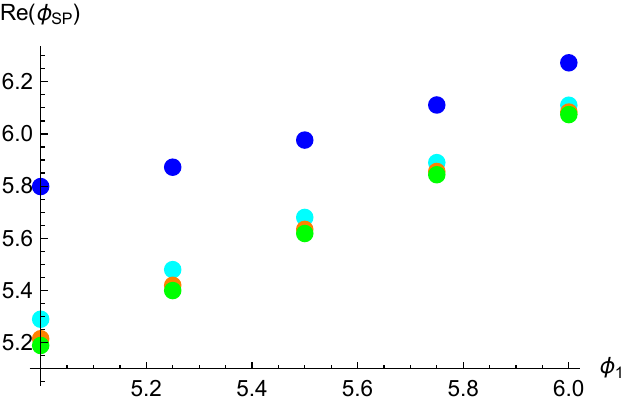}\,
	\includegraphics[width=0.3\textwidth]{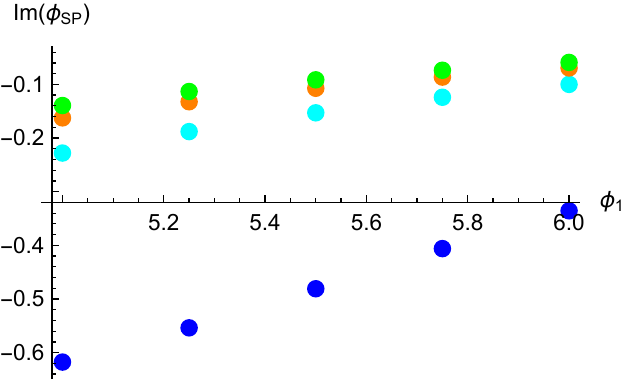}\,
	\includegraphics[width=0.3\textwidth]{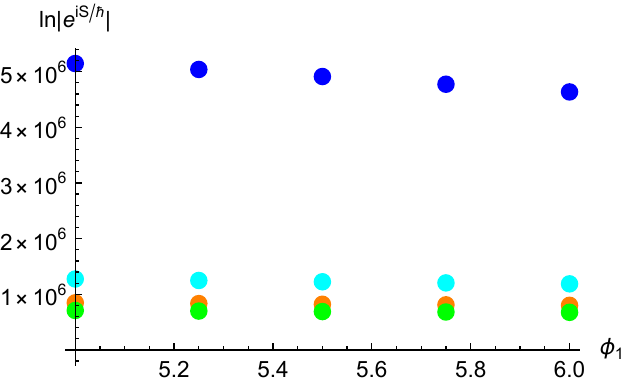}
	\caption{Optimised South Pole values and weightings for different final scalar field values $\phi_1,$ for fixed final scale factor $a_1=200,$ for different values of the internal flux $n_4.$ Smaller potentials are seen to be favoured, which is a characteristic feature of the no-boundary proposal. }
	\label{fig:ex8optimised}
\end{figure}

For different values of the flux, we obtain different potentials. At small values of the flux, the valley floor resides at negative values of the potential. For larger values of the flux, the valley floor rises and eventually merges with the top of the adjacent hill. Beyond this critical value for the flux, no minimum exists anymore, and the potential just falls off towards zero as $\chi \to \infty.$ In the latter case, there can be inflationary solutions with the field space trajectory moving to large $\chi$ values (instead of rolling along the $\phi$ direction), corresponding to a decompactification of the internal manifold. Thus, there is only a certain range of $n_4$ which gives rise to suitable potentials, keeping $\tilde\alpha=1$ fixed: when $n_4 \lessapprox 11.7,$ the valley floor drops to negative values, and since inflation cannot then occur, no-boundary solutions do not exist. Around $n_4 \approx 15.1$ the valley floor merges with the adjacent ridge in the potential, so that no extremum remains and only decompactifying solutions exist. Thus, imposing the final condition of having a large classical $4-$dimensional universe with a stable internal manifold, selects a range of fluxes. In the present model this range is
\begin{align}
    11.7 \lessapprox n_4 \lessapprox 15.1 \quad (\tilde\alpha=1)\,.
\end{align}
Within this range, no-boundary solutions can be found. We illustrate the shape of the potential valleys for the cases $n_4=12,13,14,15$ in Fig.\ \ref{fig:ex8potentials}. Larger flux corresponds to a higher potential. Fig.\ \ref{fig:ex8optimised} then shows the optimised South Pole scalar field values for no-boundary solutions with final vales $a_1=200$ and different $\phi_1,$ for the different values of the flux (the colour coding being the same as in Fig.\ \ref{fig:ex8potentials}). As one can see from the right panel in the figure, solutions that evolve at smaller values of the potential are preferred. This means that typically the no-boundary proposal will only lead to a short period of accelerated expansion, if no other conditions are imposed (though it must be sufficiently long for the inflationary attractor to get a chance to operate -- for very short periods of inflation, no no-boundary solutions can be found). We do not see this as a serious drawback at this stage, as our model is highly simplified. A realistic model would have to incorporate many more features of our universe, and would likely contain many more consistency requirements. \\

From our point of view, the main conclusion to be drawn is that the no-boundary proposal provides a mechanism for creating universes with stable internal manifolds, containing quantised fluxes. Moreover, one obtains a probability distribution not just for different solutions but also for different fluxes. Conditioned on the requirement of obtaining a large universe with a stable internal manifold, this probability distribution is non-zero only in a restricted range. Within that range, values of the flux that lead to lower potentials are preferred. Given that the fluxes are quantised, one may thus conclude that the no-boundary proposal selects the lowest quantised value that still leads to a potential containing an inflationary valley. While we can evidently not yet develop this setting into a fully realistic model of the universe  (containing, as mentioned above, realistic particle physics, etc.), we note however that this is consistent with the current non-observation of primordial gravitational waves, which would have been expected to arise in models with higher potentials.\\

\begin{figure}[h]
	\centering
	\includegraphics[width=0.8\textwidth]{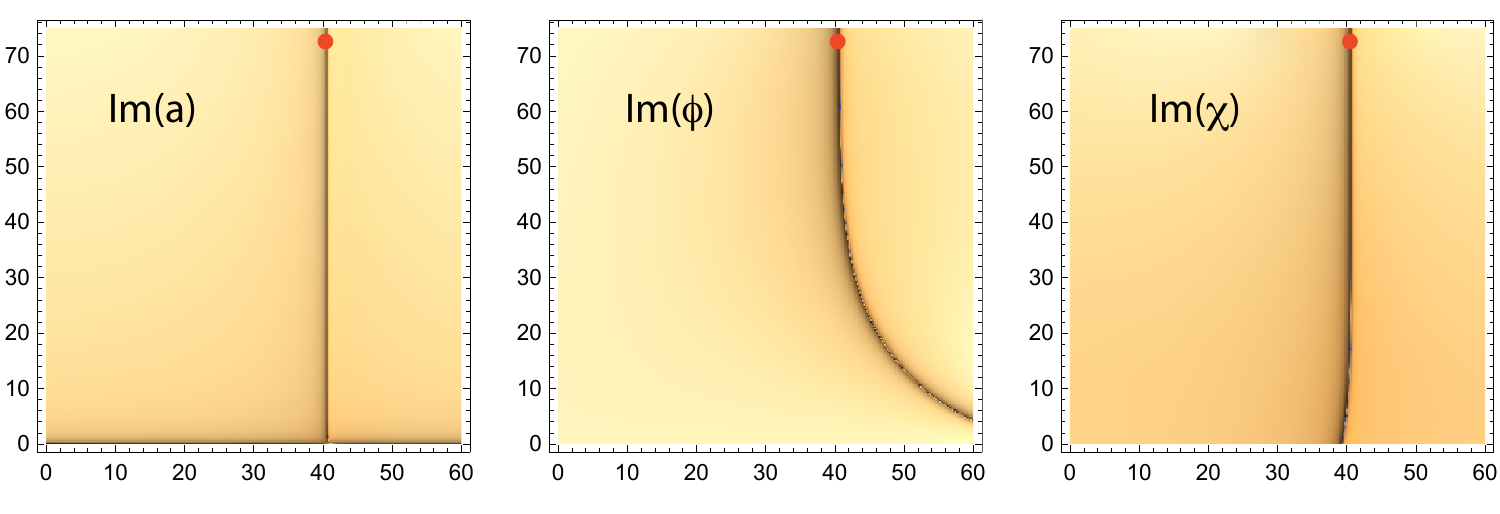}
	\caption{Imaginary field values for $a,\phi,\chi$ respectively (from left to right) in the complexified time plane (Euclidean time is in the horizontal direction, Lorentzian time vertical), for a solution in which the sphere radion starts off on a ridge in the potential and then rolls down to the valley. Dark lines correspond to zero imaginary part, \ie to the locus of real field values. The corresponding field evolutions along a Hawking type path are shown in Fig.\ \ref{fig:ex8d3}. The parameters chosen are  $\tilde\alpha=1, n_4=13.$ The solution reaches the values $a_1=200, \phi_1=6, \chi_1=2.8$ by starting from the optimised South Pole values $\phi_{SP}=6.0440-0.025359i, \chi_{SP}=3.6273 -0.065048i$ and ending at time  $\tau=40.677+73.123i,$ marked by a red dot.}
	\label{fig:ex8d3complextime}
\end{figure}

\begin{figure}[h]
	\centering
	\includegraphics[width=0.32\textwidth]{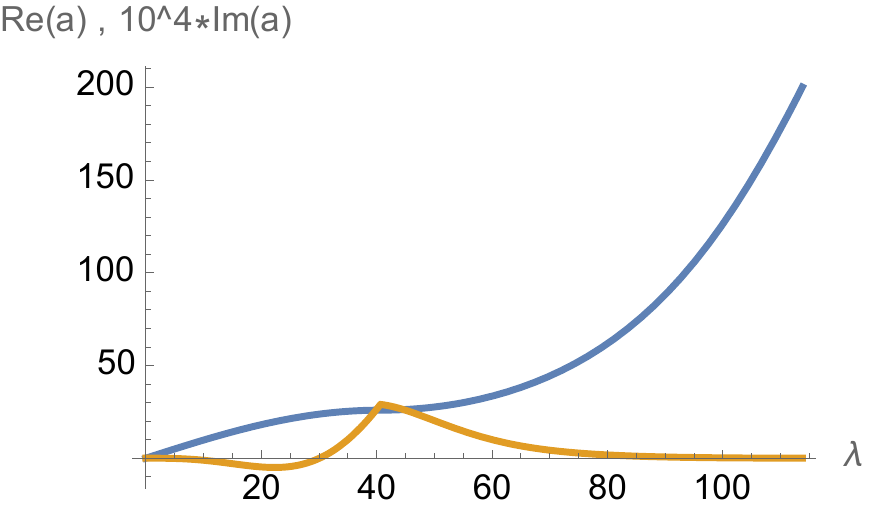}\,
	\includegraphics[width=0.32\textwidth]{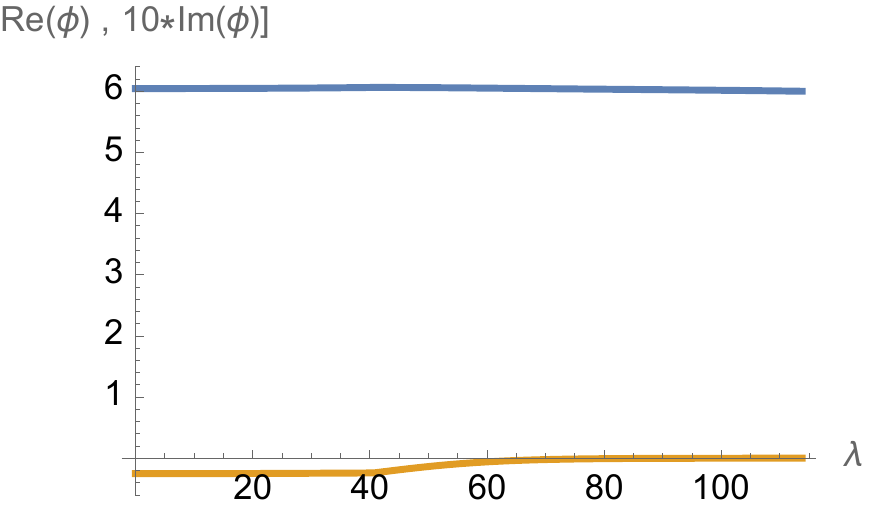}\,
	\includegraphics[width=0.32\textwidth]{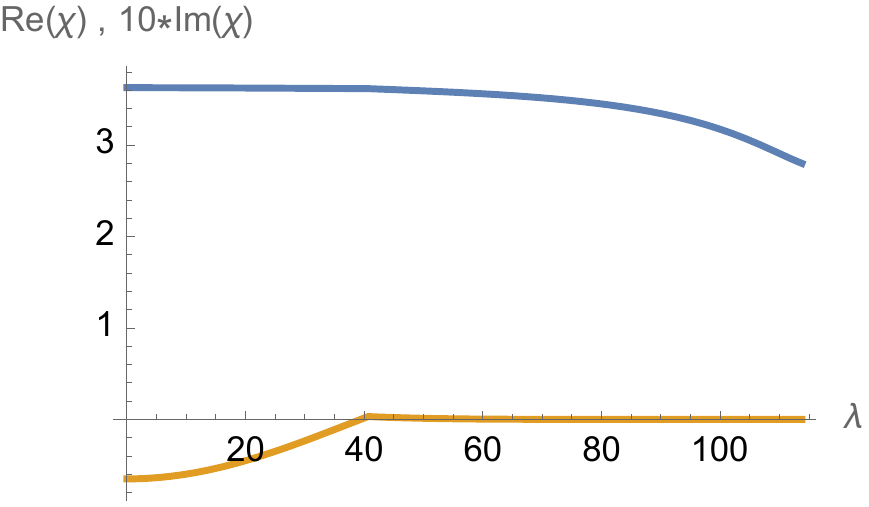}
	\caption{Field evolutions for the no-boundary solution shown in Fig.\ \ref{fig:ex8d3complextime}, along a path that starts from the South Pole in the Euclidean time direction, and then proceeds along the Lorentzian direction at $\tau=40.677$. Imaginary parts have been multiplied by $10^4$ for the scale factor and $10$ for the scalars, so as to improve visibility. For this solution, the radion $\chi$ starts out along a ridge in the potential ($\chi \approx 3.6$), cf.\ the right panel of Fig.\ \ref{fig:8dpot}, and then slides down towards the valley floor at $\chi_{min} \approx 2.8$.}
	\label{fig:ex8d3}
\end{figure}

For completeness, let us mention that further solutions can be found, though they turn out to be less probable than the ones discussed above. In particular, adjacent to the potential valleys just discussed are ridges in the potential, cf.\ again Fig.\ \ref{fig:8dpot}. A phase of accelerated expansion can take place on a ridge, with the fields rolling down towards the valley floor afterwards. Such a solution is shown in Figs.\ \ref{fig:ex8d3complextime} and \ref{fig:ex8d3}. Its action is given by $|Im(S)|=6.96 \times 10^5,$ significantly smaller than the solutions that evolve along the valley floor, cf.\ the right panel in Fig.\ \ref{fig:series}. This provides a good illustration of the way in which the no-boundary proposal is a theory of initial conditions, as it assigns different probabilities to different histories of the universe.

\section{Generalised Salam-Sezgin model} \label{sec:6d}

We also consider the example of $6-$dimensional Salam-Sezgin theory \cite{Salam:1984cj}. This theory is special from the supergravity point of view, as it contains a positive scalar potential. It may thus eventually play a role in realistic cosmological model building. However, as we will see, taken at face value the potential is not quite flat enough to allow for accelerated solutions. Still, only a small modification is required to render the potential sufficiently flat. \\

The action of the six-dimensional Salam-Sezgin theory is \cite{Salam:1984cj}
\begin{equation}
S = \int d^6x\,\sqrt{-g}\left( \frac{R}{2} - \frac{1}{2}\nabla^\mu \phi \nabla_\mu \phi  - \frac{1}{4\cdot 2!}e^{\phi}F_{(2)}^2 - \frac{1}{4\cdot 3!}e^{2\phi} H_{(3)}^2 - 4g^2e^{-\phi} \right) \,.
\end{equation}
The fluxes obey the Bianchi identities
\begin{equation}\label{bianchi}
dF_{(2)} = 0 \,,\quad dH_{(3)} = \frac{1}{2}F_{(2)}\wedge F_{(2)} \,,
\end{equation}
which can be integrated to give
\begin{equation}
F_{(2)} = dA_{(1)} \,,\quad H_{(3)} = dB_{(2)} + \frac{1}{2}A_{(1)}\wedge F_{(2)} \,.
\end{equation}
The field equations are given by equations \eqref{SSscalar} - \eqref{SSmetric} in Appendix \ref{apA}. \\

Consider the ansatz
\begin{equation}\label{ansatz}
\begin{split}
&ds^2 = -N(t)^2dt^2  + a(t)^2 d\Sigma_3^2 + b(t)^2 d\Sigma^2_2\,,\quad \phi = \phi(t) \,,\\
&F_{(2)} = f_0\vol(\Sigma_2) \,,\quad H_{(3)} = h_0\vol(\Sigma_3) \,,
\end{split}
\end{equation}
where $d\Sigma^2_{2,3}$ are the respective metrics on the spaces $\Sigma_2, \Sigma_3$ (denoted collectively $\Sigma_{2,3}$) with associated volume forms $\vol(\Sigma_{2,3})$, and $f_0$ and $h_0$ are constants. We will assume that the spaces $\Sigma_{2,3}$ are Einstein with $\text{Ric}(\Sigma_{2,3}) = \lambda_{2,3}g(\Sigma_{2,3})$, where, without loss of generality, $\lambda_{2,3} = -1, 0 ,1$. For $\lambda_{2,3} = 1$, Myers's theorem states that $\Sigma_{2,3}$ are necessarily compact, assuming that they do not have a boundary. When $\lambda_{2,3} = -1, 0$, we take $\Sigma_{2,3} = \mb{H}^{2,3}/\Gamma_{\mb{H}}, \mb{R}^{2,3}/\Gamma_{\mb{R}}$ respectively, where $\Gamma_{\mb{H}}$ and $\Gamma_{\mb{R}}$ are discrete subgroups. This ansatz satisfies the Bianchi identities in \eqref{bianchi}. \\

If we quantise the Salam-Sezgin theory, we expect the constants $f_0$ and $h_0$ to be quantised due to the quantisation of the Page charges
\begin{align}\label{quantisation}
&Q_2 = \frac{1}{2\pi}\int_{C_2}F_{(2)}\in\mb{Z} \,,\\
&Q_3 = \frac{1}{2\pi}\int_{C_3}\left(H_{(3)}-\frac{1}{2}A_{(1)}\wedge F_{(2)}\right)\in\mb{Z} \,,
\end{align}
where $C_{2,3}$ are non-trivial two and three-cycles. For our ansatz, this yields
\begin{align}
&f_0 = \frac{2\pi Q_2}{\Vol(\Sigma_2)}\,,\\
&h_0 = \frac{2\pi Q_3}{\Vol(\Sigma_3)}\,.
\end{align}
For convenience, we write down the Hodge duals of the fluxes
\begin{equation}
\begin{split}
{\ast}F_{(2)} = \frac{f_0NB^3}{A^2}dt\wedge\vol(\Sigma_3) \,,\quad {\ast}H_{(3)} = -\frac{h_0NA^2}{B^3}dt\wedge\vol(\Sigma_2) \,.
\end{split}
\end{equation}
From this, we observe that the flux equations of motion are trivially satisfied. The dynamics are thus encoded in the scalar and Einstein equations. \\

Before we get into these, let's see how one recovers the $\mb{R}^{1,3}\times S^2$ vacuum solution from \eqref{ansatz}. This vacuum does not support a three-form flux, so we set $h_0 = 0$. The scalar is stationary, so $\phi(t) = \text{constant}$. Furthermore, the geometry of the vacuum is a direct product (as opposed to a warped product), so $b(t) = \text{constant}$, and $\Sigma_2 = S^2$ ($\lambda_2 =1$). This leaves us with $N(t)$ and $a(t)$ as well as $\Sigma_3$. It can be seen that the metric
\begin{equation}
ds^2_4 = -N(t)^2dt^2 + a(t)^2d\Sigma_3^2
\end{equation}
for Einstein spaces $\Sigma_3$ is flat only if $a(t) = t$ and $\Sigma_3 = \mb{H}^3$ ($\lambda_3 = -1$) in the gauge $N(t) = 1$. Here, we have chosen the $\Sigma_3$ to be the entire hyperbolic space rather than a compact quotient to encode the entire topological structure of $\mb{R}^{1,3}$. \\

As we just reviewed, with a static internal $2-$sphere, the Salam-Sezgin solution admits a Minkowski vacuum in a Kaluza-Klein setting. This suggests that it might be possible to obtain cosmologically relevant solutions when the sphere is modified, in particular when it is allowed to be time dependent. As before, we will be chiefly interested in no-boundary solutions. For these to exist, a smooth rounding-off of the geometry must be admissible. This requires that we take $\Sigma_3=S^3,$ which we will assume henceforth. Furthermore, a regular solution must be able to satisfy the constraint
\begin{align}
  0 &= 3\left(\frac{\dot{a}}{a}\right)^2+\left(\frac{\dot{b}}{b}\right)^2 + 6 \frac{\dot{a}\dot{b}}{ab} -\frac{1}{2}\dot\phi^2+ \frac{3}{2a^2} + \frac{1}{b^2}-4g^2e^{-\phi} -\frac{f_0^2}{4b^4}e^{\phi}-\frac{h_0^2}{4a^6}e^{2\phi}\,.
\end{align} 
By inspection one can see that the last term is problematic when $a\to 0,$ as there is no other term that could cancel the associated blow-up (if $e^\phi \propto a^3$ then the potential will blow up instead). Hence we cannot have flux on the $3-$sphere, only on the internal $2-$sphere. \\

However, even with $h_0=0$ we still cannot obtain no-boundary solutions unless the dynamics also allows for inflation. With a potential of exponential form, the asymptotic solutions are well known \cite{Halliwell:1986ja}, and if the potential has the functional form $V(\phi) = e^{-c\phi}$ then the scale factor will grow as $a(t)\propto t^{1/c^2}.$ Thus one obtains accelerated expansion as long as $|c|<1.$ This is also the condition for having a dynamical attractor. Hence we can see that the Salam-Sezgin theory, with $c=1,$ resides just on the boundary between accelerated and decelerated solutions. This means that when remaining strictly in the Salam-Sezgin theory, it will not be possible to obtain no-boundary solutions \cite{Halliwell:1986bs}. We are thus led to consider a small modification of the Salam-Sezgin theory, allowing for shallower potentials: we will take the potential to be given by $V(\phi)= 4 g^2 e^{-c\phi}$. Analogously, we will generalise the coupling of the dilaton to the 2-form flux to be given by an $e^{c\phi}$ factor. Then the equations of motion read
\begin{align}
    0 &= \ddot\phi + \left(3 \frac{\dot{a}}{a} + \frac{\dot{b}}{b} \right)\dot\phi -4c g^2 e^{-c\phi} + \frac{f_0^2 c}{4b^4}e^{c\phi}\,, \label{6d1}\\
    0 &= \frac{\ddot{a}}{a} + 2 \left( \frac{\dot{a}}{a} + \frac{\dot{b}}{b}\right)\frac{\dot{a}}{a} + \frac{1}{a^2} - 2g^2e^{-c\phi} + \frac{f_0^2}{8b^4}e^{c\phi}\,, \label{6d2}\\
    0 &= \frac{\ddot{b}}{b}+ \left( 3\frac{\dot{a}}{a} + \frac{\dot{b}}{b}\right)\frac{\dot{b}}{b} + \frac{1}{b^2} - 2g^2e^{-c\phi} - \frac{3 f_0^2}{8b^4}e^{c\phi}\,, \label{6d3}
\end{align}
while the constraint reads
\begin{align}
    0 &= 3\left(\frac{\dot{a}}{a}\right)^2+\left(\frac{\dot{b}}{b}\right)^2 + 6 \frac{\dot{a}\dot{b}}{ab} -\frac{1}{2}\dot\phi^2+ \frac{3}{2a^2} + \frac{1}{b^2}-4g^2e^{-c\phi} -\frac{f_0^2}{4b^4}e^{c\phi}\,. \label{6d4}
\end{align}
Here $a$ denotes the scale factor of the intended $3$ large dimensions, $b$ is the scale factor of the internal $S^2$, and $\phi$ is the dilaton. 

\begin{figure}[h!]
	\centering
	\includegraphics[width=0.8\textwidth]{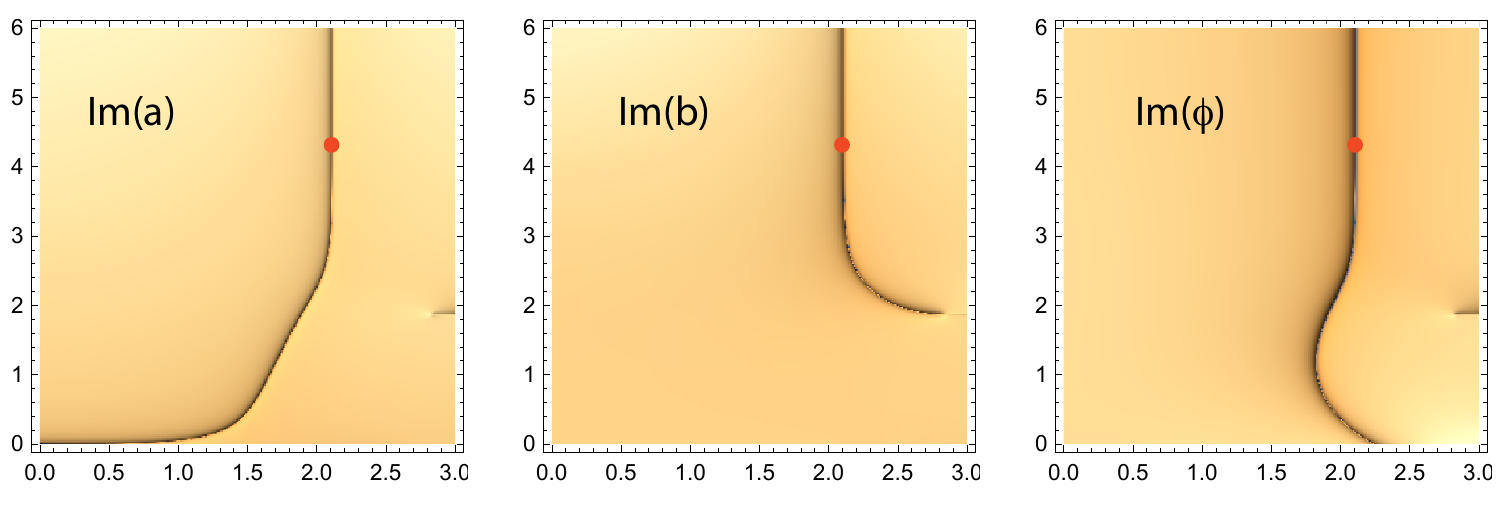}
	\caption{This graph shows a density plot of the imaginary parts of $a, b$ and $\phi$ in the complexified time plane. Darker shades mean smaller imaginary part, so that the black lines mark the locus of real field values. At late times, the dark lines become aligned with the Lorentzian time direction and start overlapping, hence a classical universe is obtained. For this example we used $c=1/10, f_0=1$ and the final values $(\phi_1,a_1,b_1)=(1,15,2)$ are reached at time $\tau=2.1135+4.3388i.$ The optimised parameters are $\phi_{SP}=1.5621  -0.29160i, b_{SP}=0.54365 + 0.29203i.$}
	\label{fig:nbex1}
\end{figure}

\begin{figure}[h]
	\centering
	\includegraphics[width=0.3\textwidth]{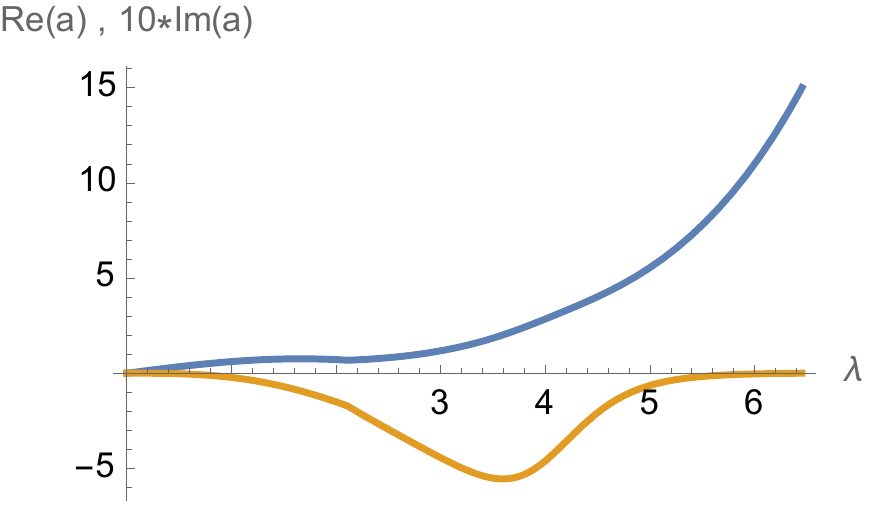}
	\includegraphics[width=0.3\textwidth]{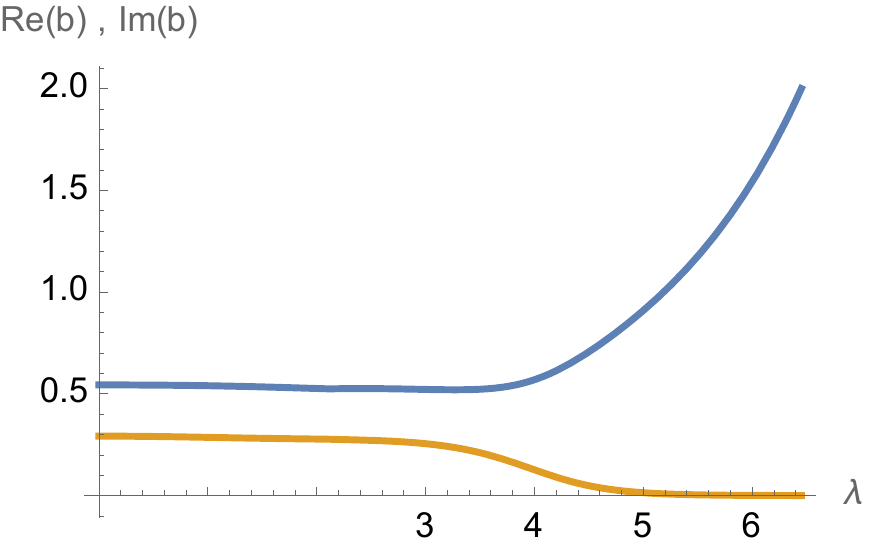}
	\includegraphics[width=0.3\textwidth]{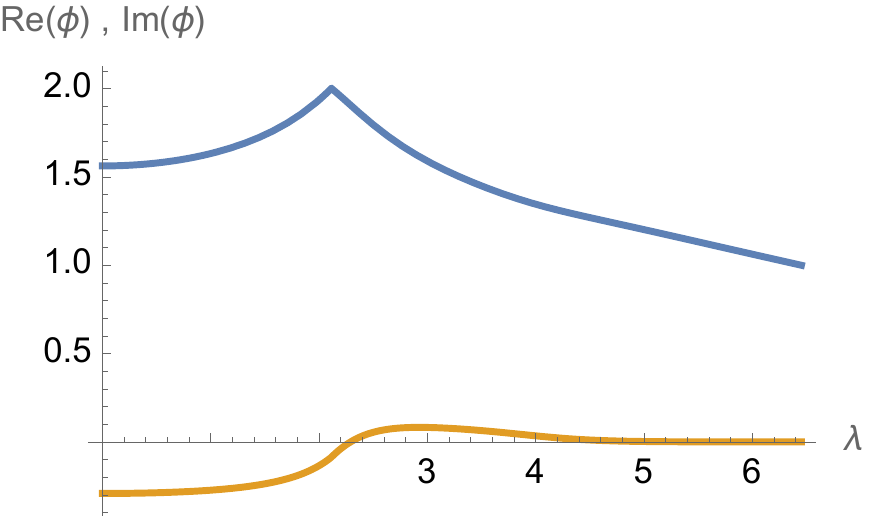}
	\caption{These graphs show the field evolutions along a path in the complexified time plane that, starting from the South Pole of the solution, first follows the Euclidean time direction and then the Lorentzian time direction, for the solution shown in Fig.\ \ref{fig:nbex1}. Blue curves stand for real parts, orange curves for the imaginary parts of the fields (the imaginary part of the scale factor $a$ has been enhanced for better visibility). The internal scale factor $b$ keeps expanding, indicating that this solution will decompactify.}
	\label{fig:nbex2}
\end{figure}

We are looking for no-boundary solutions, \ie for solutions that are regular at the origin and reach desired real values $(\phi_1,a_1,b_1)$ on the final $S^3 \times S^2$ hypersurface. Regularity at the origin implies that the geometry must become Euclidean there. More precisely, if we Taylor expand solutions starting at $a(t=0)=0,$ then we find the following series expansions (in  Euclidean time $\tau = it$ ), 
\begin{align}
    \phi(\tau) =&  \phi_{SP} +  \left(\frac{c e^{c \phi_{SP} }}{32 b_{SP}^4}-\frac{1}{2} c e^{-c \phi_{SP} }\right) \tau ^2 + {\cal O}(\tau^4)\,, \\
    a(\tau) =&  \frac{1}{\sqrt{2}}\tau + \left( \frac{5 e^{c \phi_{SP} }}{288 \sqrt{2} b_{SP}^4}-\frac{1}{36 \sqrt{2} b_{SP}^2}-\frac{e^{-c \phi_{SP} }}{18 \sqrt{2}}\right)  \tau^3 + {\cal O}(\tau^5)\,, \\
    b(\tau) =& b_{SP} + \left( -\frac{3 e^{c \phi_{SP} }}{64 b_{SP}^3}-\frac{1}{4} b_{SP} e^{-c \phi_{SP} }+\frac{1}{8 b_{SP}}\right) \tau^2 + {\cal O}(\tau^4)\,.
\end{align}
Here $\phi_{SP}$ and $b_{SP}$ are arbitrary (and in general complex) integration constants we optimise using a Newtonian algorithm, as described in Section \ref{sec:8d}. An example of a solution is given in Figs.\ \ref{fig:nbex1} and \ref{fig:nbex2}. As seen from Fig.\ \ref{fig:nbex2}, the scalar field is initially complex but it quickly becomes real valued. Once it is real, it simply rolls down the exponential potential, as expected. The scale factor $a$ becomes real rather fast, and then expands in an accelerated fashion. The ``internal'' scale factor $b$ takes longer to become real, but then it eventually starts expanding too. This is because the scalar acts as a source for the scale factor of the sphere, and since $\phi$ keeps rolling down its potential, the scale factor $b$ has to keep evolving too. Thus we obtain a large universe with flux, but unfortunately the extra dimensions keep growing, so that we do not obtain a realistic universe.\\

Future work may allow for a resolution of this issue. There are several avenues that seem worth exploring: one could try to stabilise the dilaton too, so that both $b$ and $\phi$ would remain constant. The equations of motion \eqref{6d1}--\eqref{6d4} suggest that this is possible if $e^{c\phi}=8g^2b^2,$ with $f_0g=\frac{1}{2}.$ This looks promising, except that the ``large'' scale factor is forced to remain Euclidean, as the equations reduce to $\dot{a}^2=-\frac{1}{2}.$  Perhaps this obstacle can be overcome by adding further matter fields, that can act as a source for the scale factor. Another possibility would be to investigate more general compactifications, also from $10$ or $11$ dimensions. The Salam-Sezgin model in any case possesses an interesting embedding in higher dimensions \cite{Cvetic:2003xr} containing a $3-$dimensional hyperbolic internal space, with remarkable gravity localising properties \cite{Crampton:2014hia,Erickson:2021psj}. This setting thus seems worthwhile of further exploration.

\section{Discussion} \label{sec:discussion}

A central question of theoretical physics is which aspects of our world arose out of necessity, and which out of happenstance. Quantum gravity, and string theory in particular, has greatly expanded the scope of this question. On the one hand, there are stringent consistency conditions, that for instance demand that extra spatial dimensions exist. On the other hand, string theory has provided a setting in which a single theory leads to a myriad of solutions, so that in the end we are faced with the question as to why the particular solution that describes our universe was given preference over all the others. This question may largely turn out to be one of cosmology. \\

\begin{figure}[h!]
	\centering
	\includegraphics[width=0.7\textwidth]{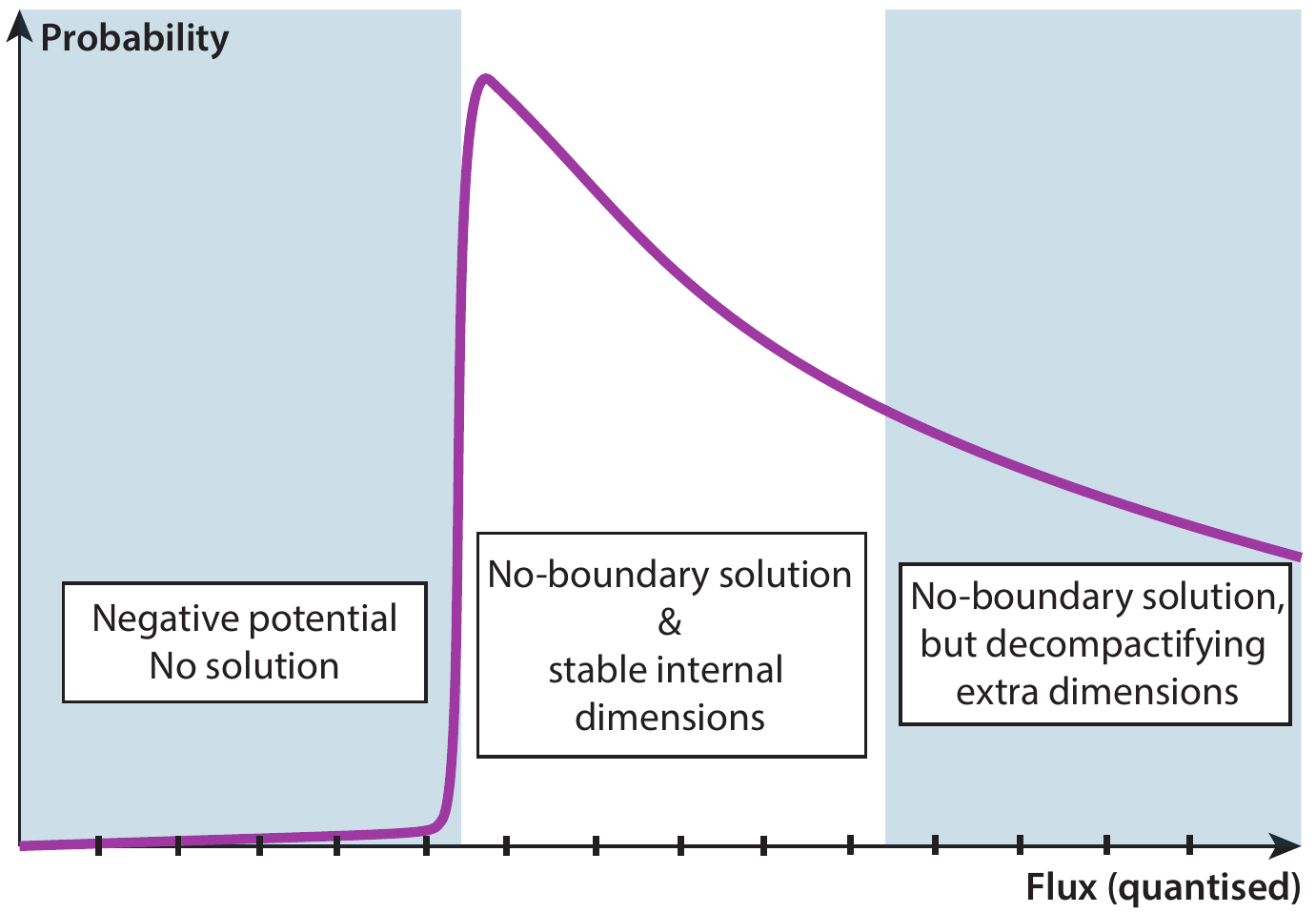}
	\caption{A schematic summary, based on the model of Section \ref{sec:8d}: no-boundary solutions only exist when a dynamical attractor is present, in this case inflation. But the potential is positive and inflationary only for sufficiently large flux values. When the flux is too large, the inflationary valley in the potential disappears and the extra dimensions decompactify. No-boundary solutions with stable extra dimensions only exist for a range of flux values, and in this way the no-boundary proposal restricts the possible amounts of flux. In this manner, cosmology can act as a selection principle on the landscape of string theory solutions.}
	\label{fig:summary}
\end{figure}

Our work illustrates how cosmology may act as a strong selection principle on solutions. In the context of string inspired toy models, we have described solutions that explain how a universe with internal dimensions, and moreover with internal flux, can arise. These solutions obey no-boundary initial conditions, and as such they have finite curvature everywhere (and finite action), and hence they are fully trustworthy already at the semi-classical level \cite{Jonas:2021xkx}. As we have discussed, these solutions are only possible if the effective scalar potential allows for an inflationary region, which only occurs for certain ranges of the internal flux. Since internal flux is quantised, at most a handful of values of flux are typically expected to be viable. It is in this sense that cosmology acts as a selection principle on solutions. In addition, the no-boundary proposal provides probabilities for different amounts of flux, and for different solutions, see also our summarising Fig. \ref{fig:summary}.\\

Quite generally, internal flux enhances the scalar potential. Since the no-boundary proposal assigns higher probability to lower potential values, smaller amounts of flux come out as preferred, as long as they are compatible with a large, classical universe. In general, string solutions contain many moduli, and one has to explain why all (or at least most) of them are stabilised, especially once an initial inflationary phase comes to an end. Here we would simply remark that the preference for lower values of the potential may help in this regard, as it naturally puts all moduli at the minima of their potentials. Then it remains to explain why just one (or perhaps a few) started out higher up on the potential, leading to an inflationary phase. Explaining this aspect might be feasible in the sense that classical universes are only seen to arise if a phase of accelerated expansion (or perhaps one of high-pressure contraction \cite{Battarra:2014xoa}) have taken place. \\ 

At present our models are not sufficiently sophisticated to address detailed questions about our universe. For this, further ingredients are evidently necessary, such as fields that describe particle physics, dark matter and a late quintessence phase. Also, we did not attempt to embed our models into complete string theoretic solutions. These are obvious and desirable goals for future research. Rather, we have demonstrated a general principle, namely that cosmology drastically restricts the landscape of quantum gravity solutions. This seems a promising direction to pursue.

\newpage

\section{Acknowledgments}

We would like to thank Sergio H\"{o}rtner for discussions. JLL gratefully acknowledges the support of the European Research Council in the form of the ERC Consolidator Grant CoG 772295 ``Qosmology''. The work of KSS was supported in part by the STFC under Consolidated Grant ST/P000762/1.

\begin{appendix}

\section{Higher-dimensional equations of motion} \label{apA}

\subsection{$R^4$ theory in 8 dimensions}

The field equations of the $R^4$ theory after the field redefinition in \eqref{conformal} for $D=8$ and $p=4$ are 
\begin{align}
&d{\ast}d\phi - V'(\phi)\vol_8 = 0 \,, \label{8dscalar}\\
&d{\ast}F_{(4)} = 0 \,,\\
&R_{MN} = \partial_M\phi\partial_N\phi + \frac{1}{3}V(\phi)g_{MN} + \frac{q^2}{12}\left(F^{}_{MP_1P_2P_3}F_N^{\ph{N}P_1P_2P_3} - \frac{1}{8}F_{(4)}^2g_{MN}\right) \,,\label{8dmetric}
\end{align}
where 
\begin{align}
    V(\phi) = \tilde\alpha\left(1 - e^{-\sqrt{\frac{6}{7}}\phi} \right)^{\frac{4}{3}}\,,
\end{align}
and $\vol_8 = {\ast}1$.\footnote{Our convention for the Hodge dual on a $D$-dimensional Lorentzian manifold in terms of an orthonormal frame $\{e^m\}$, is $\ast(e^{m_1}\wedge\cdots\wedge e^{m_p}) = \frac{1}{q!}\epsilon_{\ph{m_1\cdots m_p}n_1\cdots n_q}^{m_1\cdots m_p} e^{n_1}\wedge\cdots\wedge e^{n_q}$, where $\epsilon_{012\dots(D-1)}=1$, $e^{012\dots(D-1)}= -1$, and $q = D-p$.} 

\subsection{Salam-Sezgin theory}

The field equations for the Salam-Sezgin theory are
\begin{align}
&d{\ast}d\phi - \frac{1}{4}e^{\phi}F_{(2)}\wedge{\ast}F_{(2)} - \frac{1}{2}e^{2\phi} H_{(3)}\wedge{\ast}H_{(3)} + 4g^2e^{-\phi}\vol_6 = 0 \,,\label{SSscalar}\\
&d(e^{\phi}{\ast}F_{(2)}) + e^{2\phi} F_{(2)}\wedge{\ast}H_{(3)} = 0 \,,\\
&d(e^{2\phi}{\ast}H_{(3)}) = 0 \,,\\
&R_{MN} = \partial_M\phi\partial_N\phi + 2g^2e^{-\phi}g_{MN} + \frac{1}{2}e^{\phi}\left(F^{}_{MP}F_N^{\ph{N}P} - \frac{1}{8}(F_{(2)})^2g_{MN}\right) \nonumber \\
&\qquad\quad + \frac{1}{4}e^{2\phi}\left(H^{}_{MPQ}H_N^{\ph{N}PQ} - \frac{1}{6}(H_{(3)})^2g_{MN}\right) \,,\label{SSmetric}
\end{align}
where $\vol_6 ={\ast}1$.

\end{appendix}

\bibliographystyle{utphys}
\bibliography{bibNBSS.bib}

\end{document}